# Examining the variability of cloud hydrometeors in the context of Indian summer monsoon rainfall predictability

Ushnanshu Dutta[1,2], Anupam Hazra[1*], Subodh Kumar Saha[1], Hemantkumar S. Chaudhari[1], Samir Pokhrel[1], and Mahen Konwar[1]

[1]Indian Institute of Tropical Meteorology, Ministry of Earth Sciences, India

[2]Department of Atmospheric and Space Sciences, Savitribai Phule Pune University, Pune, India

**Key finding:**

1) The sub-seasonal variability of cloud ice (CI) and cloud water (CW) are linked with seasonal Indian summer monsoon rainfall (ISMR).

2) Variability of cloud condensates on synoptic and super-synoptic bands has strong association with global predictors (e.g. ENSO).

3) Improved simulation of CI leads to capture their observed global teleconnection.

**Corresponding author:**

Dr. Anupam Hazra, hazra@tropmet.res.in

# Abstract

Skilful prediction of the seasonal Indian summer monsoon (ISM) rainfall (ISMR) at least one season in advance has great socio-economic value. It represents a lifeline for about a sixth of the world's population. The ISMR prediction remained a challenging problem with the sub-critical skills of the dynamical models attributable to limited understanding of the interaction among clouds, convection, and circulation. The variability of cloud hydrometeors (cloud ice and cloud water) in different time scales (3-7 days, 10-20 days and 30-60 days bands) are examined from re-analysis data during Indian summer monsoon (ISM). Here, we also show that the 'internal' variability of cloud hydrometeors (particularly cloud ice) associated with the ISM sub-seasonal (synoptic + intra-seasonal) fluctuations is partly predictable as they are found to be tied with slowly varying forcing (e.g., El Niño and Southern Oscillation). The representation of deep convective clouds, which involve ice phase processes in a coupled climate model, strongly modulates ISMR variability in association with global predictors. The results from the two sensitivity simulations using coupled global climate model (CGCM) are provided to demonstrate the importance of the cloud hydrometeors on ISM rainfall predictability. Therefore, this study provides a scientific basis for improving the simulation of the seasonal ISMR by improving the physical processes of the cloud on a sub-seasonal time scale and motivating further research in this direction.

## 1. Introduction:

The interannual and intraseasonal variability of ISMR controls the livelihood and health of over two billion people and regulates the country's economy through food production (Gadgil, 2007). The prediction of this variability well in advance can help farmers in crop management and government to take preparation for natural calamities (e.g., flood and drought) associated with them. There are many studies on ISMR prediction and its predictability (e.g., Charney and Shukla, 1981, Shukla, 1981, Kang and Shukla, 2006, Lau and Yang, 1996, Sperber et al. 2000, Wang et al., 2008, Wang, 2005; Webster et al., 1998, Goswami et al. 2006). However, the seasonal prediction of the ISMR has remained a significant challenge due to the skill of the most current generation of climate models being sub-optimal (Krishna Kumar et al. 2005; Rajeevan et al. 2012).

The ISM is a coupled climate system (Webster et al., 1998), which is strongly influenced by subseasonal variability (synoptic to intraseasonal scales) (Webster et al., 1998; Goswami and Xavier 2005, Goswami et al., 2003). Contrary to these, Krishnamurthy and Shukla (2007, 2008) have argued that the monsoon intraseasonal oscillations (MISOs) do not contribute significantly to the seasonal mean ISMR and the contribution of synoptic rainfall on the year-to-year variation of ISMR is also questionable (Krishnamurthy and Ajayamohan 2010). However, low-pressure systems (LPS), which have a typical length scale of 3-5 days (i.e., synoptic systems) are known as effective rain-bearing mechanisms (e.g., Mooley, 1973; Sikka, 1977; Yoon & Chen, 2005). Furthermore, the high-frequency synoptic components, which were considered as "climate noise," and therefore unpredictable, are also found to be partly predictable, as they are tied with slowly varying forcing, e.g., El Niño and Southern Oscillation (ENSO) (Saha et al. 2019). While the sub-seasonal components of ISMR are partially predictable, it is not clear whether the variability of different cloud condensates [i.e., cloud ice (CI), and cloud water (CW)] are also linked with the slowly varying predictable component. Waliser et al. (2009) have shown that cloud ice is an important and major challenge in most global climate models. It is also known that cloud microphysical processes play a significant role in modulating the intraseasonal variability of ISM (e.g., Bony et al. 2015; De et al. 2016; Kumar et al., 2017; Hazra et al., 2017a, 2020; Dutta et al. 2020). Bony et al. (2015) have reported that clouds are essential for climate sensitivity studies in climate models. The misrepresentation of clouds, precipitation, and circulation has been continued for many new-generation models. It is emphasized by De et al. (2019) that the interaction between cloud and large-scale circulation remains a grey area of climate science. But the variability of cloud hydrometeors (cloud ice and cloud water) in different time scales (3-7 days, 10-20 days and 30-60 days bands) are not attempted so far during Indian summer monsoon (ISM).

The role of clouds in general and cloud condensates like CI and CW, in particular, is essential for the simulation of synoptic-scale disturbances by AOGCMs (Atmosphere-Ocean General Circulation Models) with higher fidelity. This is also a critical requirement for the success of seasonal prediction of south Asian monsoon rainfall in particular and tropics in general. Nevertheless, the improved microphysical scheme in coupled forecast system version 2 (CFSv2: Saha et al., 2014) increases the seasonal prediction skill of the ISMR beyond the potential limit of predictability (based on signal-to-noise-ratio framework; Saha et al., 2019). Is that improvement in skill by chance? Or the role of cloud condensate to link regional convection with large-scale circulation through a reasonable simulation of the space-time characteristics of the sub-seasonal components is consistent with observations?

In the present endeavour, we attempted to address the following questions: -

i) What is variability of cloud hydrometeors (cloud ice and cloud water) in different time scales (3-7 days; 10-20 days and 30-60 days bands) during ISM?

ii) Whether the sub-seasonal variance of cloud condensate of different types is predictable on a seasonal time scale?

iii) What is the intraseasonal relationship between cloud condensates, convection, and circulation?

iv) Whether improved ISMR prediction skill is associated with the improved teleconnection of cloud condensate with global predictors?

The present work is organized as follows: Section 2 discusses the data used and the methodology adopted. Results obtained from observational and reanalysis data are discussed in section 3. Section 4 discusses results obtained from sensitivity experiments using a coupled climate model. Finally, the major findings of the study are summarized in section 5.

## 2. Data and Methodology:

Modern-Era Retrospective analysis for Research and Applications (MERRA) (Rienecker et al., 2011) datasets (daily averaged) are used for zonal winds and vertical profile of mixing ratio of CI and of CW. Monthly mean observed rainfall datasets are taken from the Global Precipitation Climatology Project (GPCP) (Adler et al., 2003). Outgoing Longwave Radiation (OLR) data are from the National Oceanic and Atmospheric Administration (NOAA) (Liebemann and Smith, 1996). The Hadley Centre Global Sea Ice and Sea Surface Temperature (HadISST) data (Rayner et al., 2003) are also used in this study. All these data are considered for a common period of

1981-2010. Seventeen years of daily rainfall data (1998-2014) are taken from Tropical Rainfall Measuring Mission (TRMM) (TRMM-3B42, Huffman et al., 2007). The same is used for comparison of the rainfall climatology with GPCP data. GPCP daily rainfall data (1997-2010) is also used to find the composite of synoptic events.

To find the relationship between the sub-seasonal variance of CI and CW, with the ISMR and ENSO, we adapted the method from Saha et al., (2019) study. The CI and CW are vertically averaged over pressure levels followed by spatial averaging (land region only) over all India (70ºE - 90ºE, 10ºN - 30ºN) and central India (72ºE - 88ºE, 18ºN - 28ºN). This yields daily time series (30 years; 1981-2010) of CI and CW for all India and central India. Then we have calculated the CI and CW time-series (for each harmonic) after filtering out the harmonics up to 150th (i.e., 365/150 = 2.4 days periodicity), one by one by Fourier analysis. The first four harmonics (i.e., 0, 1, 2, 3) together represent the annual cycle. The remaining harmonics together represent the total anomaly of periodicity 2-91.25 days band (in a 365 days calendar year, the fifth harmonics represents 365/4=91.25 days periodicity). One by one, harmonics are removed up to 150 (corresponds to 365/150=2.43 days) and reconstructed back into the time series of daily CI and CW anomaly. Therefore, for each year, there will be 147-time series (i.e., an anomaly with harmonics greater than 3, 4, 5, ..., 150 corresponds to the anomaly of 2-91.25, 2-75, ...., 2-2.43 days band respectively). Then the seasonal (JJAS) variance of each filtered time series (i.e., 147-time series) is calculated.

Unlike in a band-pass filter, these variances will show their association with ISMR or any predictors in a continuous band of frequency. Since the sub-seasonal components are considered the building blocks of the ISMR, the variances of cumulative bands will also assess a model's performance more holistically. The internal variance of each sub-seasonal harmonics is computed for each June-September (JJAS) season, and the same is correlated with the seasonal (JJAS) mean of ISMR (70ºE - 90ºE, 10ºN - 30ºN, the land region only) averaged rainfall and Niño 3.4 (170ºW - 120ºW, 5ºS - 5ºN) SST.

The 30-years daily anomaly of central India averaged cloud condensate (CI and CW) is computed after removing the first four harmonics (i.e., annual cycle) from the daily data, which is considered as an unfiltered (raw) anomaly in this study. The anomaly is then passed by a suitable Lanczos filter for computing synoptic (i.e., the period between 3-7 days), Quasi-Biweekly Mode (QBM, i.e., the period between 10-20 days), and MISO (i.e., period between 30-60 days) anomaly. We selected only the JJAS period (122 days) from all these anomaly data (raw, synoptic, QBM, and MISO) and constructed a time-series, consisting of now 122 days in each year for each set of anomaly data. Next, we calculate the variance for each year (i.e., the internal variance of 122 days for each year) to evaluate the 'internal variance' for raw, synoptic, QBM, and MISO scale.

The climate models are used to understand complex earth systems and test hypotheses using long free climate simulation, which is more than 30 years. Here we have performed 50 years of climate simulation using coupled forecast system version 2 (CFSv2) for control and modified experiments. The details of model simulations for control and modification are available in Hazra et al. (2017a, 2020). We have also used the last 30 years of data from 'control' and modification experiments ('MPHY') for understanding the role of the variability of CI on ISMR. We have considered cloud condensates below the temperature level of -15ºC as purely CI for model simulations (Zhao and Carr, 1997).

## 3. Results:

### 3.1 Mean state and variability of cloud hydrometeors (cloud ice and water) and rainfall during Monsoon

Clouds play a seminal role in governing rainfall variability (e.g., Chaudhari et al., 2016) through modulation of heating (Hazra et al., 2017b; Hong and Liu, 2015) and induced circulation (Kumar et al. 2014). Chaudhari et al. (2016) showed the significant correlation of level-wise (high, mid, and low) cloud fraction with rainfall over the ISMR region. The JJAS climatology of rainfall from GPCP (1981-2010), TRMM (1998-2014), and CI, CW (1981-2010) from MERRA is shown in Figure 1 over the global tropics. In global tropics (30S:30N), it is evident that mean CI distribution is better associated with the mean rainfall distribution (pattern correlation = 0.76) than that of CW (pattern correlation = 0.59); demonstrating the importance of cloud ice during the JJAS over India (Hazra et al. 2017b). The satellite observations also revealed that about 40–50% of rainfall events originate from ice melting (Field and Heymsfield 2015). Moreover, over the Indian land region, CI and CW are co-located with maxima of rainfall during the monsoon season. There is little rain in the east Pacific, which is in line with the marginal CI there. The same is also true for the Southern tropical Indian Ocean. The rainfall over the mid equatorial pacific is from the low-level clouds mainly consists of CW (Fig. 1). The variability of cloud hydrometeors (cloud ice and cloud water) in different time scales (3-7 days, 10-20 days and 30-60 days band) are shown (Fig. 2) from re-analysis data during Indian summer monsoon (ISM). It is also interesting to note that variances of cloud ice (Fig. 2a-c) in all three bands are strong as compared to cloud water (Fig. 2d-f). The synoptic variance (3-7 days) of CI and CW are more followed by QBM (10-20 days) and MISO (30-60 days), which is similar to the rainfall variances (Fig. 3) for the same bands.

**3.2 Predictability of sub-seasonal variability of cloud condensates**

Recently, Saha et al. (2019) found that rainfall variability in the synoptic-scale has the strongest correlation with the seasonal ISMR than the other two bands of Monsoon Intraseasonal Oscillations (i.e., 10-20 days and 30-60 days). The rainfall variance in the synoptic-scale is also linked with the slowly varying predictable component (e.g., ENSO). This leads to an important question: whether the statistics of the cloud condensates on the sub-seasonal scale are also linked with seasonal JJAS mean rainfall and ENSO? This may support developing a better cloud microphysics scheme to improve the sub-seasonal simulation, which eventually may lead to an improved seasonal prediction of the ISMR. Therefore, we investigate the possible relationship of the sub-seasonal variability of CI and CW with ISMR and Niño3.4 SST in observations.

It is already known from many past studies that the maximum number of synoptic disturbances formed over the head of the Bay of Bengal (BoB) and pass across central India. It is essential to know how the synoptic variability of cloud condensates over this zone is associated with convection and SST. The synoptic variability over CI (averaged over central India) shows a strong and significant (more than 95%) relationship with the seasonal OLR over the east Pacific (NIÑO 3 and NIÑO 3.4) region (Fig. 4a), which is weak in case of CW (Fig. 4b). Correlation of synoptic variance of CI and seasonal mean OLR reveals significant EQUINOLR (index for EQUINOO) (Gadgil et al., 2019) pattern also. Interestingly, the correlation of the synoptic variability of CI with SST (Fig. 4c) reveals canonical El-Niño mode. On the other hand, CW also shows a similar pattern, but with less prominence (Fig. 4d). We have also examined the relationships described above in QBM (Fig. 5), MISO (Fig. 6), and RAW (Fig. 7). The cloud condensate (CI or CW) does not depict any significant relationship with SST over east Pacific regions like the synoptic-scale for these scales. EQUINOLR pattern also absent in these scales.

To investigate the relationship of cloud condensate with ISMR and teleconnection with El-Niño across sub-seasonal scales, we have presented the correlation of variance of cloud condensates (both CI and CW) averaged over all India and central India with seasonal mean ISMR and NIÑO 3.4 SST across the period 2.460 days (Fig. 8). We find that the variance of CI over All-India depicts the highest correlation in the 3-5 days period, consistent with the same analysis for the Central India region, i.e., the monsoon core region. The association of CI with NIÑO3.4 also shows (Fig. 8a) the strongest inverse relationship in the same period, which indicates that this synoptic variability has the potential to be predicted by ENSO. A well and above 95% significant correlation exists between CW and mean ISMR at the QBM, which is absent for CI. This highlights the role of CI in synoptic mode and CW in the QBM of monsoon oscillation (Hazra et al., 2020). The correlation of mean ISMR, as well as NIÑO3.4 SST with the variance of CW and CI, reveals inverse relationships for MISO.

## 3.3 Relationship of Cloud Condensates with Convection and Circulation on sub-seasonal scale

The low-level Jetstream (LLJ) is one of the significant components of the Asian summer monsoon (Krishnamurti and Bhalme, 1976). LLJ also shows Spatio-temporal variability on an intraseasonal scale (Goswami et al., 1998), associated with MISO. Joseph and Sijikumar (2004) found that over BoB (10ºN - 20ºN, 80ºE - 100ºE, box shown in Fig. 7d), convective heat source strengthens LLJ at 1-2 days lead. On the other hand, Chaudhari et al. (2017) have shown the relationship between OLR and high-level clouds during JJAS. Therefore, we explore the relationship of OLR and LLJ with the cloud condensates, which will help connect the missing link between clouds convection and circulation. We have presented lead-lag (-30 to 30 days window) correlation (Fig. 9) of all India (central India) averaged cloud condensates (both CI and CW) with the BoB averaged OLR (OLR BoB) and the zonal component of 850 hPa wind (U850 BoB) from their 30 years of daily JJAS data. We see that convection (i.e., less OLR) first triggers the formation of CW, followed by CI formation as convection deepens; where correlation with CW (CI) peaks at around eight (five) days lead of OLR. Now, the cloud microphysical processes generate more CW in the background of deep convection, which is represented here as the highest (negative) peak between CW and OLR-BoB at two days lead of OLR. After the zero-lag between OLR and cloud condensates, the correlation between CW and OLR decreases and becomes significantly positive after 4-5 days (CW leading).

On the other hand, CI shows the highest (negative) correlation with OLR at around 2-3 days' lead (CI leading). This implies that CI can further trigger the convection, as it continues to stay in the upper level due to its weaker fall velocity than CW. CI can trap the longwave radiation in contrast with that of CW. This leads to a further decrease in the OLR and a deepening of the convection.

The correlation of condensates with the circulation reveals that circulation significantly leads the CW, and the strength gradually increases towards its maxima, nearly at zero lag. At ten days' lead of the CW, the correlation becomes insignificant. An interesting relationship between CI and U850 BoB is observed (Fig. 9b). The correlation starts increasing at a lead (U850 BoB leading) from thirty days to eighteen days, gradually decreasing to five days. As CI leads, correlation with U850 BoB increases and peaks at around five days lead. Hence, our results reveal that convection first forms CI (i.e., upper-level clouds) that further energizes the convection, which in turn stirs up the circulation.

## 4. Results from sensitivity experiments using a coupled climate model

The sub-seasonal variance of CI and seasonal mean Niño3.4 SST from the control experiment shows a positive relationship, which is opposite to that of reanalysis (Fig. 10a). This is significantly improved in MPHY and is in line with the reanalysis. On the other hand, the control experiment shows multiple peaks for correlation of CI with mean ISMR in the synoptic-scale, which is also improved in MPHY. Interestingly, the control experiment seems to be not sensitive to monsoon's interannual variability. It shows a flat correlation of intraseasonal variability with the mean rainfall, which is also improved in the MPHY experiment (Fig. 10a).

The synoptic variance of CI condensate averaged over central India is correlated with SST of the global ocean at every grid point for the control and MPHY (Fig. 10b, c). The similarity of the correlation patterns with the canonical ENSO SST confirms the modulation of the sub-seasonal fluctuations of CI over the ISM region through ENSO teleconnection (Fig. 4c). The correlation is inversely represented in control (Fig. 10b) simulation, consistent with Figure 10a. The same is realistically captured in modification (MPHY) (Fig. 10c). It implies that the representation and simulation of deep convective clouds, which involve ice phase processes, strongly modulates ISMR prediction (e.g., Hazra et al., 2017b) and predictability, which further leads to the improvement in the skill of monsoon (e.g., Saha et al., 2019).

It is also noted that most of the lows and depressions initiate over the BoB and move towards central India (Mooley, 1973; Sikka, 1977), which contributes to about 45-55% of seasonal ISMR (Yoon & Chen, 2005). Therefore, during the depression (or any synoptic-scale activity), stronger vertical velocity will help to uplift more moisture below the freezing level to form supercooled cloud water and cloud ice and snow/graupel via several microphysical processes (e.g., condensation, deposition, freezing, riming/accretion) (Hazra et al., 2016). It further invigorates the cloud by releasing latent heat during microphysical conversions through thermo-dynamical phase-changes and effect on dynamics (Hazra et al., 2013). The large-scale background conditions may help suppress and enhance convection, which is eventually reflected in the vigor of synoptic and MISOs events. The lead-lag composite of 3 to 7 day filtered cloud ice anomaly from MERRA data is shown in Figure 11 to demonstrate the evolution of synoptic events. The maximum synoptic events are identified for major and two reference time series (i) over BoB, where maximum lows and depressions initiate/form and (ii) over central India. The maxima of the synoptic event signify by lag-0. Figure 11 demonstrates an evident westward propagation and escalation over the head BoB and central India. The rainfall (1997-2010) composites for corresponding events are overlaid to find any lead/lag between rainfall (or convection) maxima (or minima) and CI maxima (or minima). Figure 11 reveals that CI leads the rainfall events, consistent with our earlier results (Fig. 9), where we saw the CI leads the convection. The similar lead-lag composite of CI at synoptic scale from model sensitivity experiments (control, MPHY) are evaluated to determine the evolution of synoptic events (Fig. 12, 13) and their association with rainfall. The modified model (MPHY) shows better the spatiotemporal evolution (Fig. 13) of synoptic events (e.g., lows and depressions) of CI and rainfall, where CI leads the rainfall noticed as seen in observation (Fig. 11).

## 5. Summary

In the backdrop of the continued advancement in recent times in the coupled climate model, the improvement of seasonal prediction of ISMR is still sub-critical. The poor simulation of synoptic and super-synoptic variances in coupled climate models is most likely due to the poor representation of convection/microphysics parameterizations (Goswami and Goswami, 2016; Hazra et al., 2017b, 2020, Dutta et al., 2020) and maybe one of the reasons for poor 'actual skill' of ISMR (Saha et al., 2019). Saha et al. (2019) have shown that the improved 'actual skill' of the MPHY model is achieved due to the correct representation of synoptic variances of rainfall as compared to observation. However, the question was unravelled, whether the formation of cloud condensate also behaves similarly. This answer will help target the improvement in the present convection/microphysics scheme of the coupled climate model. The new understanding for improving prediction of ISMR can be summarized below:

1. The proper representation of cloud and convection in the synoptic time scale is essential for the seasonal mean ISMR. Therefore, the activity of the synoptic disturbances (the variance of cloud condensate, particularly CI) has a strong positive correlation with the seasonal mean rainfall.

2. The finding reveals that the synoptic-scale variance of cloud hydrometeors is linked with slowly varying forcing (e.g., El Niño and Southern Oscillation). Therefore, cloud microphysics (formation of all cloud hydrometeors like cloud ice, water, snow, graupel, etc.) plays a crucial role in the seasonal prediction of ISMR.

3. The lead-lag correlation of the CI with OLR-BoB reveals the role of CI in triggering the convection. The convection, in turn, strengthens the circulation.

4. The sensitivity model simulation using coupled forecast system version 2 (CFSv2) with modified microphysics demonstrates the support of the model development activity based on new observational/reanalysis findings. The lead/lag relationship between cloudice and rainfall is well captured in modified model (MPHY) as compared to control as shown in the spatio-temporal evolution.

This new horizon of monsoon prediction by improving physical processes (here, cloud, convection, and circulation as shown in schematic Fig. 14) will motivate researchers to take up the challenges for improving the ISMR skill far ahead using the new generation coupled climate model, which likely to begin a new chapter of reliable seasonal ISM forecasting.

**Acknowledgments:**

We thank MoES, Government of India, and Director IITM for all the support to carry out this work. We also acknowledge the freely available data sets used in the study, which are as follows, HadiSST (https://rda.ucar.edu/datasets/ds277.3/), MERRA cloud condensate data (https://disc.gsfc.nasa.gov/datasets/MAI3CPASM_5.2.0/summary), GPCP rainfall data (https://psl.noaa.gov/data/gridded/data.gpcp.html), NOAA OLR data (https://psl.noaa.gov/data/gridded/data.interp_OLR.html) and TRMM rainfall data (https://disc.gsfc.nasa.gov/datasets/TRMM_3B42_Daily_7/summary?keywords=TRMM_3B42_Daily_7). The coupled climate model simulated data for the sensitivity experiments are available in the link below https://drive.google.com/drive/folders/1-bBsDk1ooASTlABv0T8hcoI_BBgZGm5P?usp=sharing. We also acknowledge Mrs. Chandrima Mallick, IITM-Pune for her contribution to prepare the schematic (Figure 14).

**Figures:**

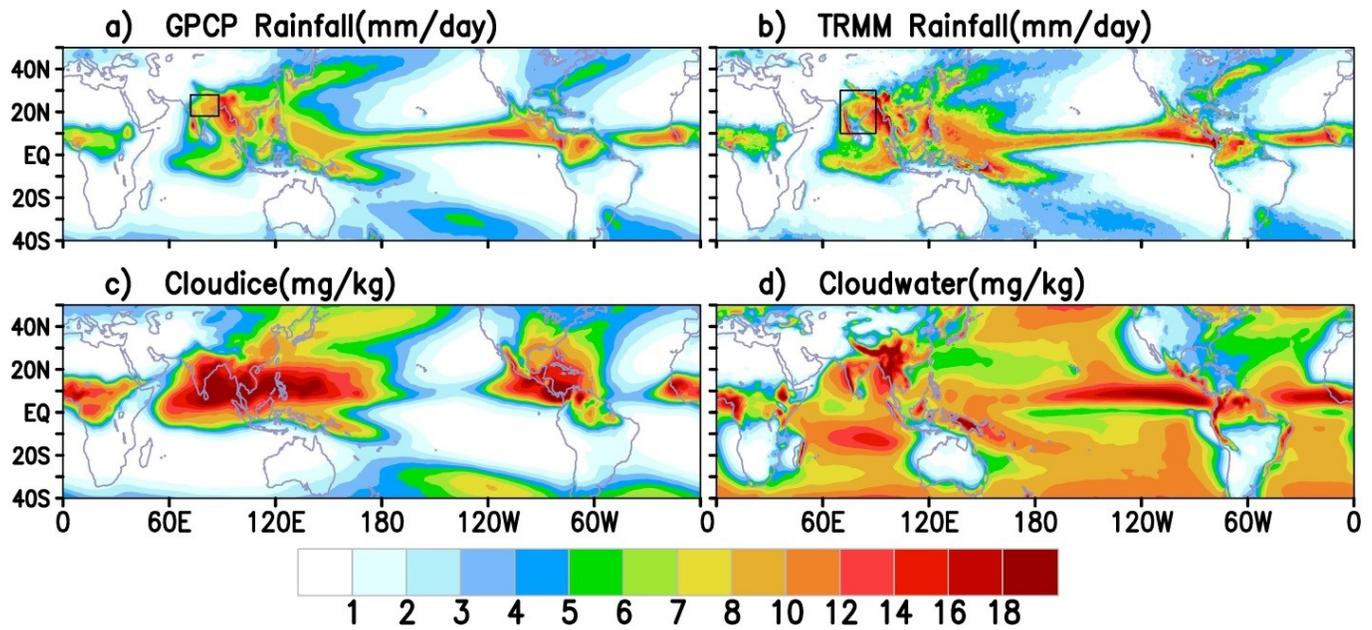

**Figure 1:** JJAS Climatology of (a) GPCP rainfall (b) TRMM rainfall (c) Cloudice and (d) Cloudwater. Except for TRMM (1998-2014) all are of 30 years (1981-2010). The box of Central-India is shown in Fig. 1a and All-India box is shown in Fig. 1b.

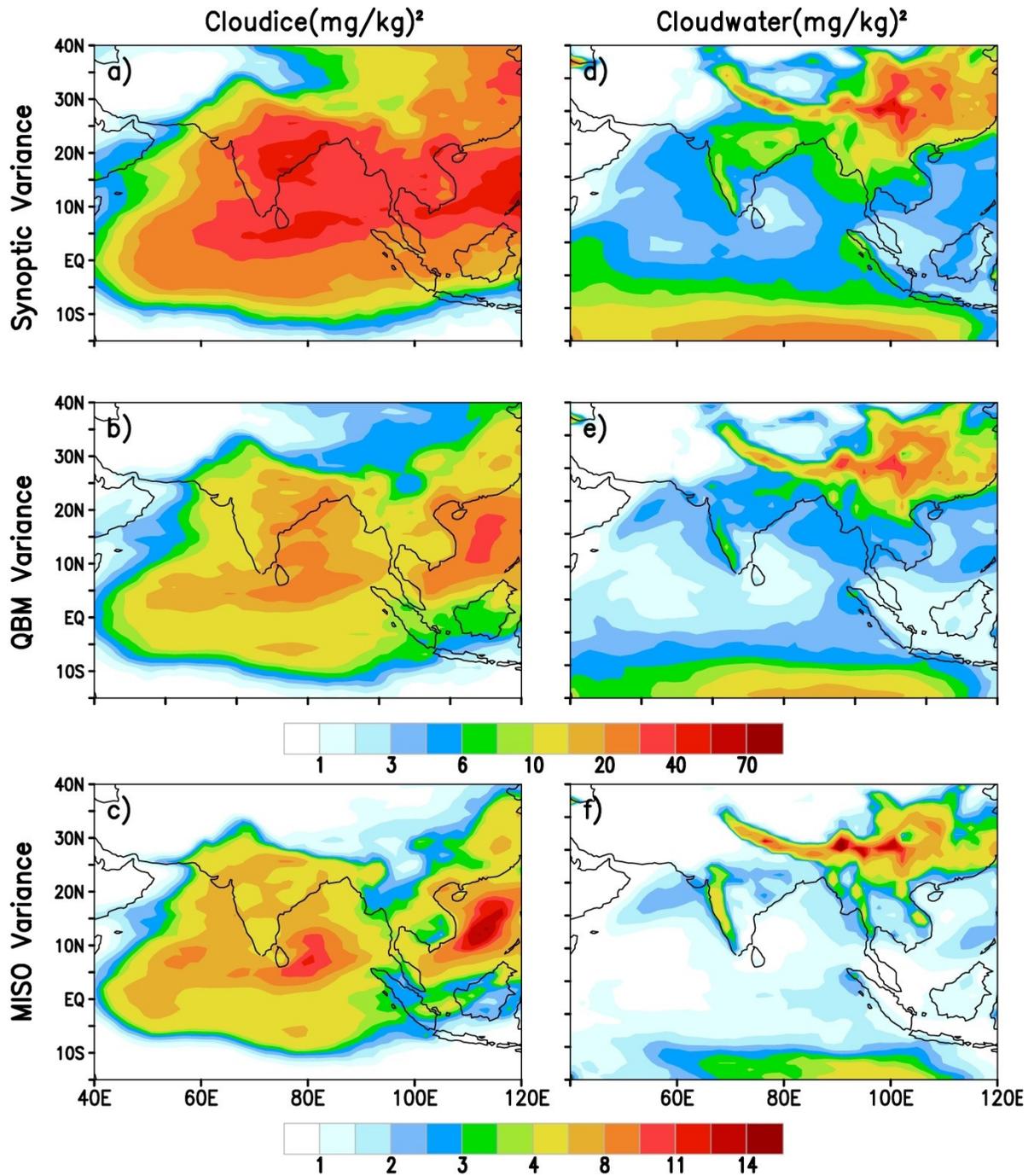

**Figure 2:** The climatology of variances in different time scales of 3-7 day (a,d: synoptic), 10-20 day (b,e: quasi bi-weekly mode, QBM) and 30-60 day (c,f: MISO) filtered cloud hydrometeors [cloud ice (a-c) and cloud water (d-f)] during the JJAS season from re-analysis (MERRA).

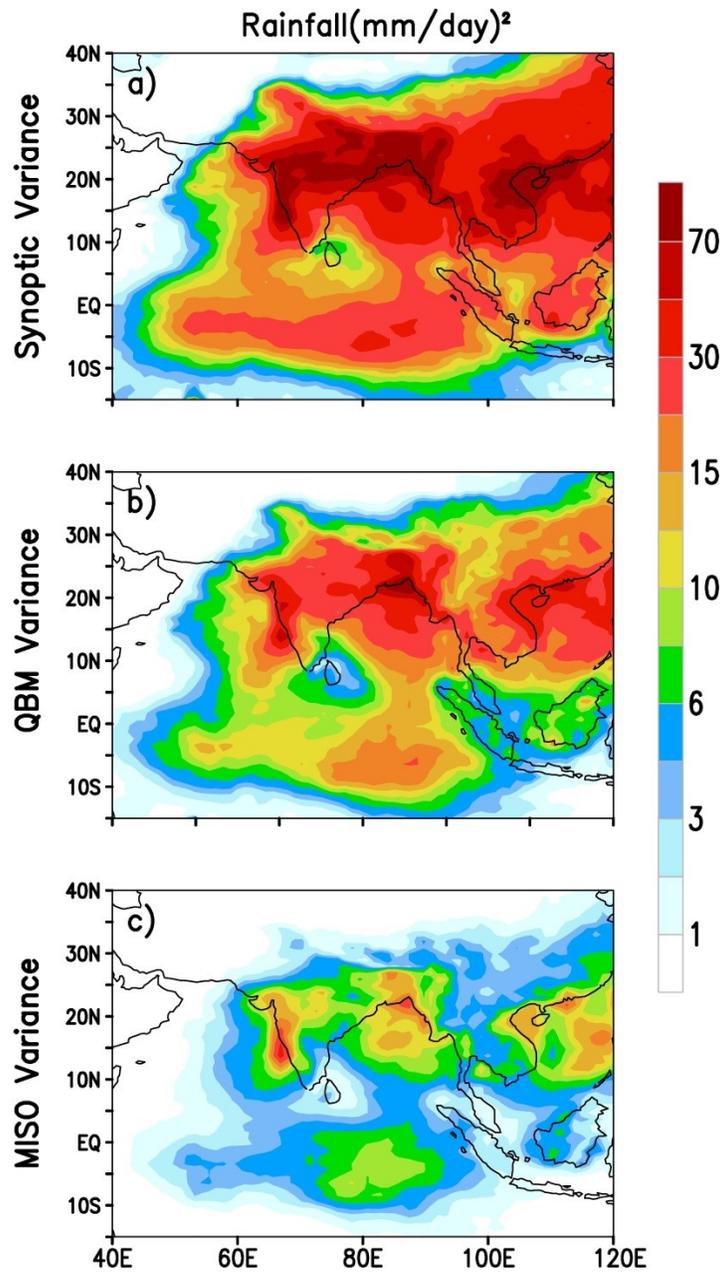

**Figure 3:** The climatology of variances in different time scales of 3-7 day (a: synoptic), 10-20 day (b: quasi bi-weekly mode, QBM) and 30-60 day (c: MISO) filtered precipitation during the JJAS season from GPCP.

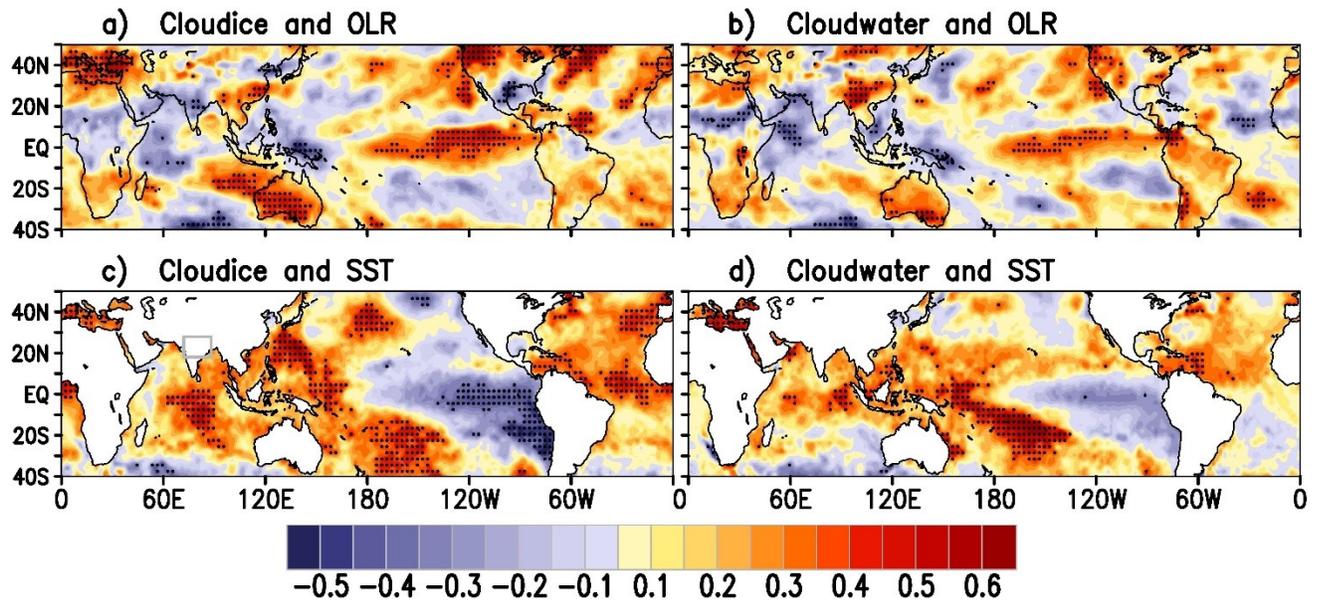

**Figure 4:** Correlation of averaged Synoptic (period 3-7 days) variance of cloud condensates (Cloudice and Cloudwater) averaged over Central India (box is shown in 2c) with Global OLR and SST. Correlation of greater than 95% significance is stippled.

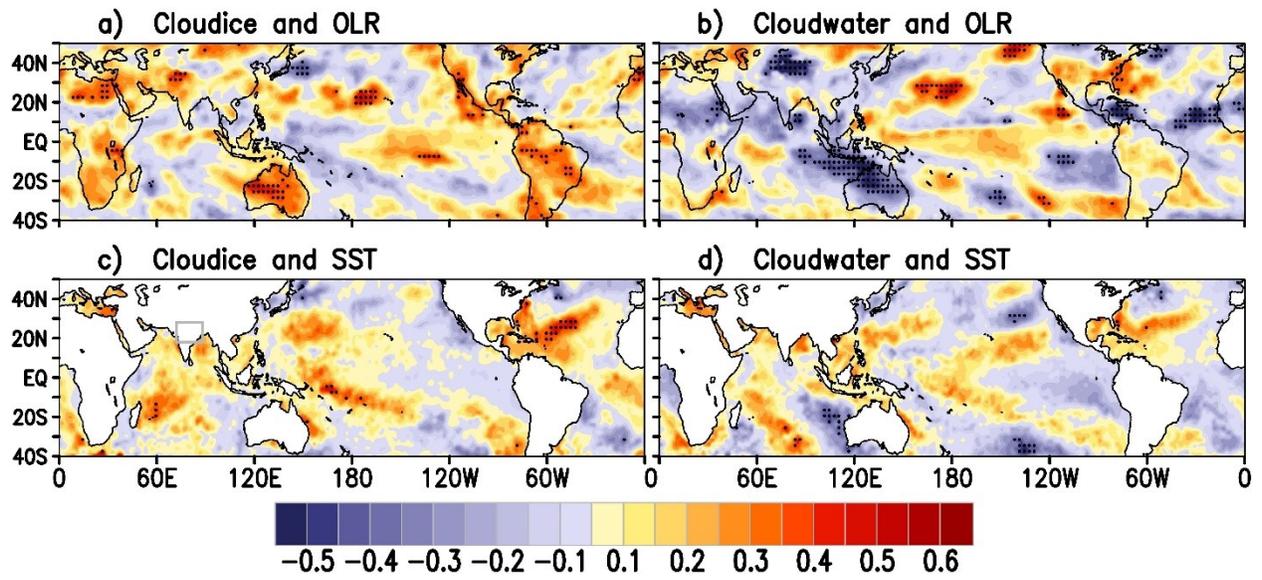

**Figure 5:** Same as Figure 4 in the paper but with averaged QBM (period 10-20 days) variance.

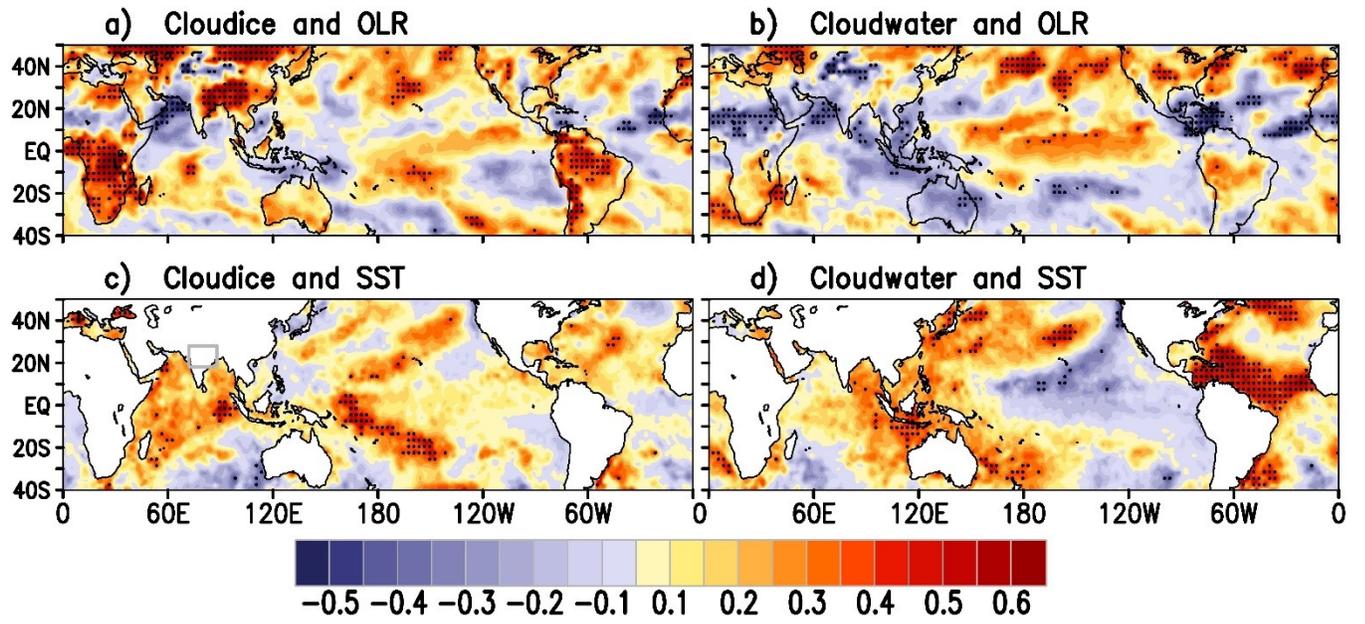

**Figure 6:** Same as Figure 4 in the paper but with averaged MISO (period 30-60 days) variance.

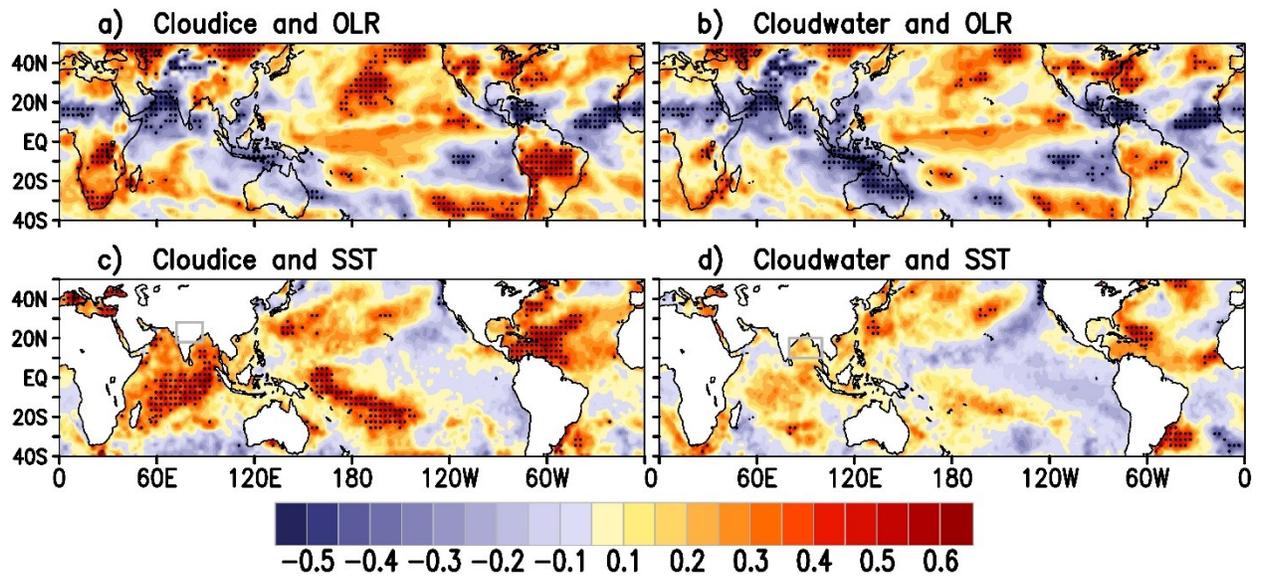

**Figure 7:** Same as Figure 4 in the paper but with averaged unfiltered (RAW) variance.

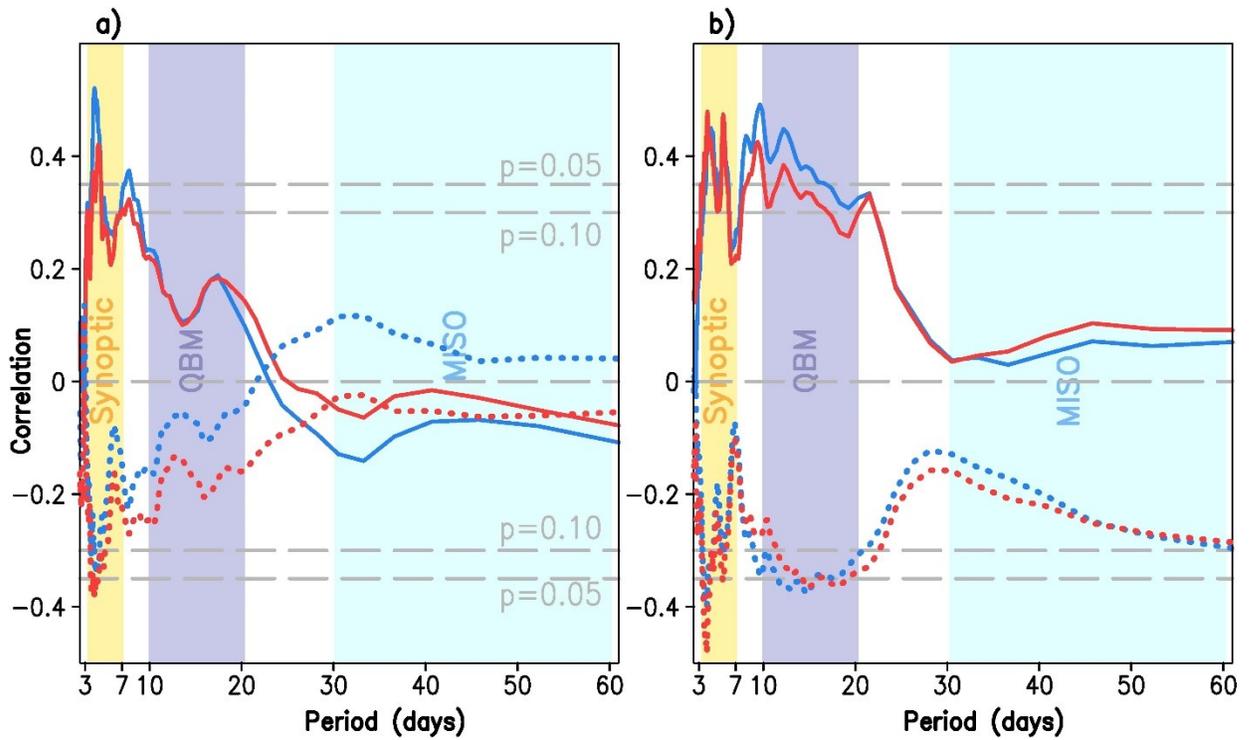

**Figure 8:** The seasonal (June to September average) mean ISMR and Niño3.4 SST anomaly correlated with the subseasonal variance of cloud-ice and cloud-water at various time bands (or period). (a) Correlation between variance of All India (Central India) averaged cloudice with JJAS mean ISMR (solid) and NINO 3.4 SST (dotted) in blue (red) lines. (b) Same as of (a) but for cloudwater. Significant correlation of 90% (p=0.10) and 95% (p=0.05) are also shown.

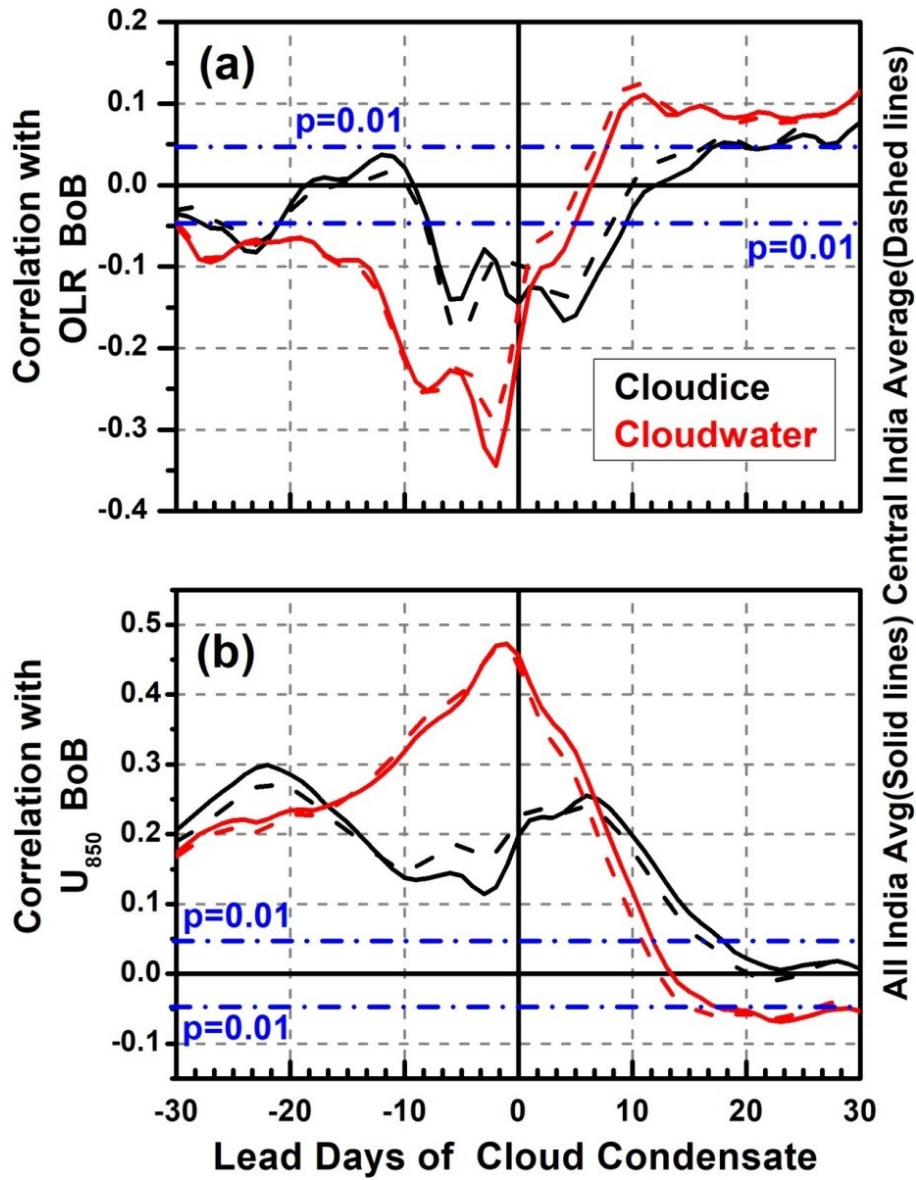

**Figure 9:** Lead-Lag Correlation of All India (Central India) averaged cloud condensates with Bay-of Bengal (BoB box is shown in Fig. 5d) averaged a) OLR and b) Horizontal wind of 850 hPa pressure level (each of JJAS daily data of 30 years). Significant correlation of 99% (p=0.01) is also shown.

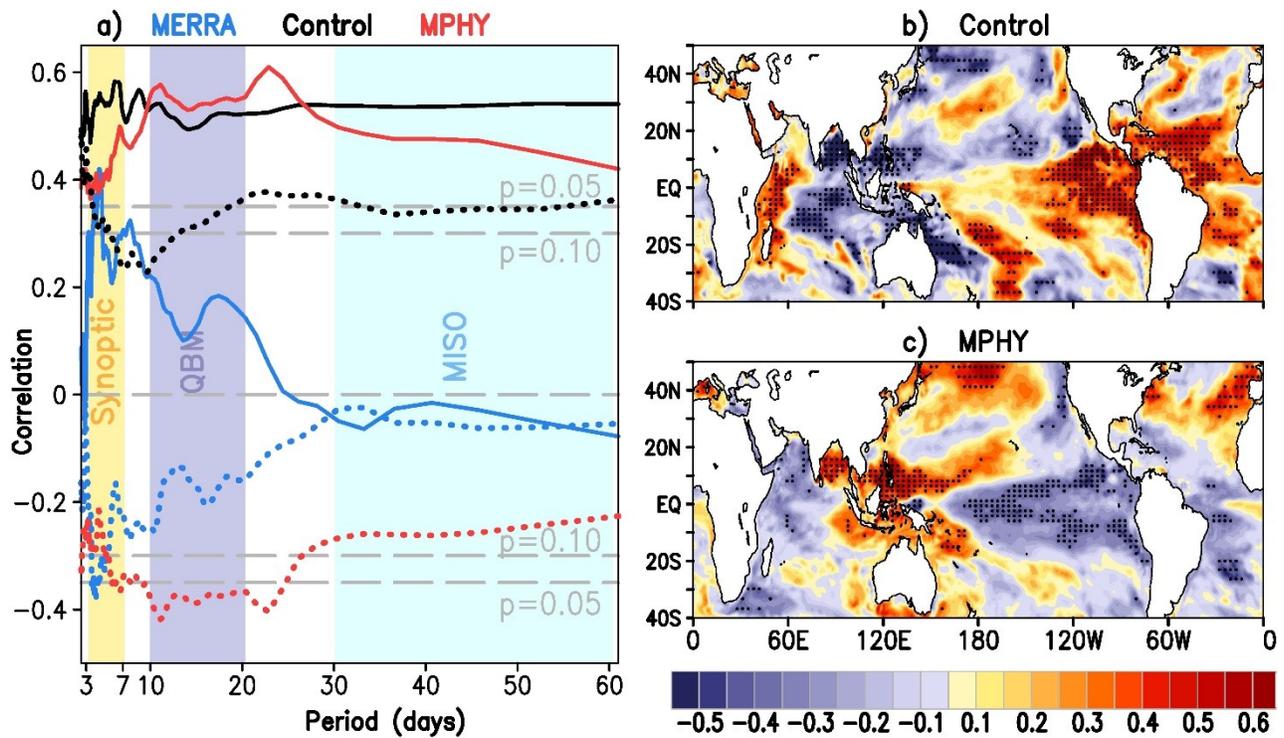

**Figure 10:** (a) Same as Figure6 (a) but for CFSv2 model experiments (Control and MPHY) and MERRA (for Central India averaged cloudice). Solid (dashed) lines of respective colors are correlation of Cloudice with JJAS mean ISMR (NINO 3.4 SST anomaly). Dashed horizontal lines marked as p=0.01 (p=0.05) shows 90% (95%) significance. b (c) are the Correlation of averaged Synoptic (period 3-7 days) variance of Cloudice averaged over Central India (box is shown in 4c) with Global SST from CFSv2 model simulation with Control (MPHY).

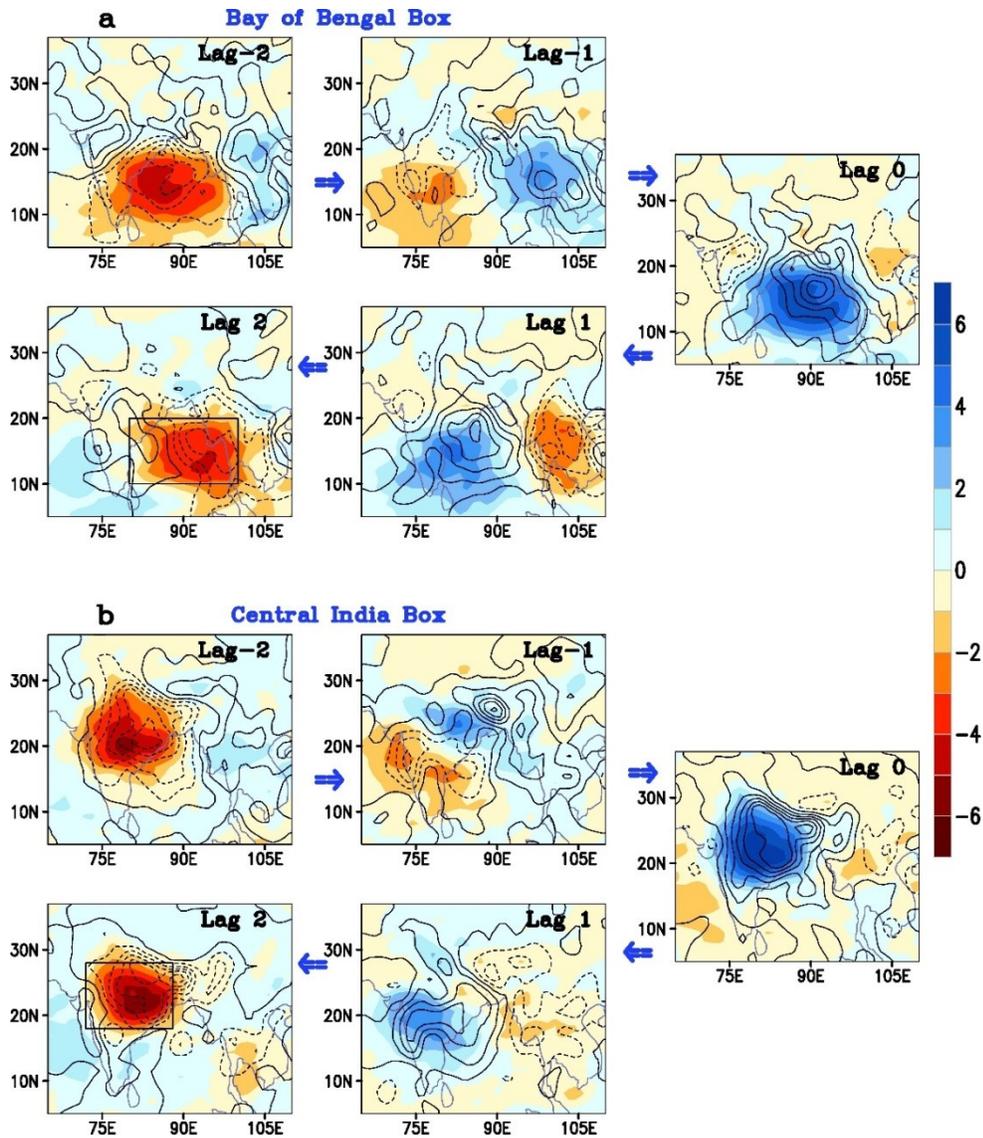

**Figure 11:** The lead-lag composite of 3- to 7-day filtered cloud-ice during June–September from MERRA. The maxima of an event (i.e., lag 0) is identified using area averaged filtered cloud-ice over a box in (a) Bay of Bengal (Lon: 80ºE - 100ºE, Lat: 10ºN - 20ºN) and (b) central India (Lon: 72ºE - 88ºE, Lat: 18ºN - 28ºN). The events (central India: 197, BoB:199) when cloud-ice greater than 1.5 standard deviations are considered for the composites. Rainfall (1997-2010) composites (Central India: 99; BoB: 73) for corresponding events are in contours.

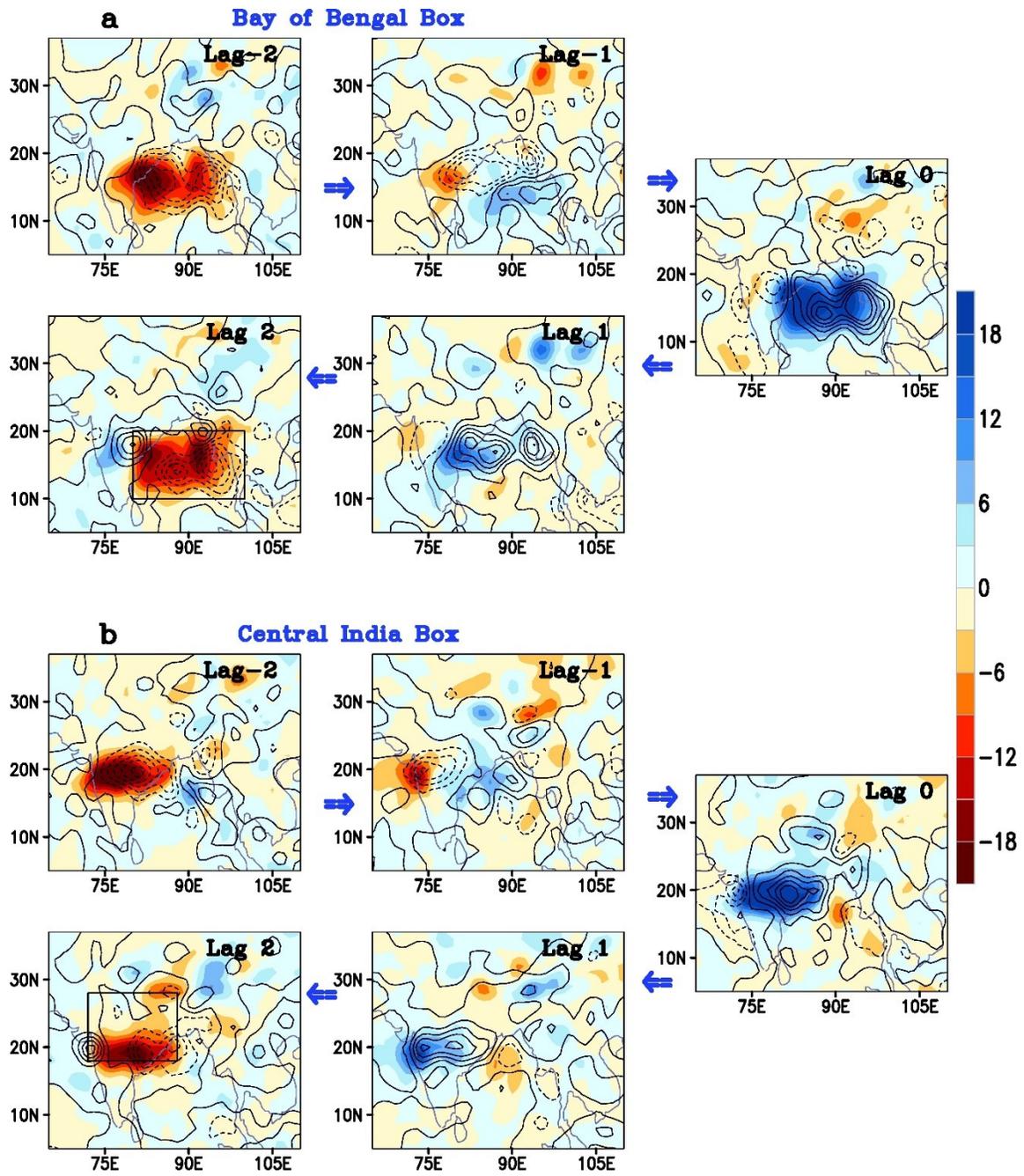

**Figure 12:** Same as Figure 11, but for for CFSv2 model experiments (Control).

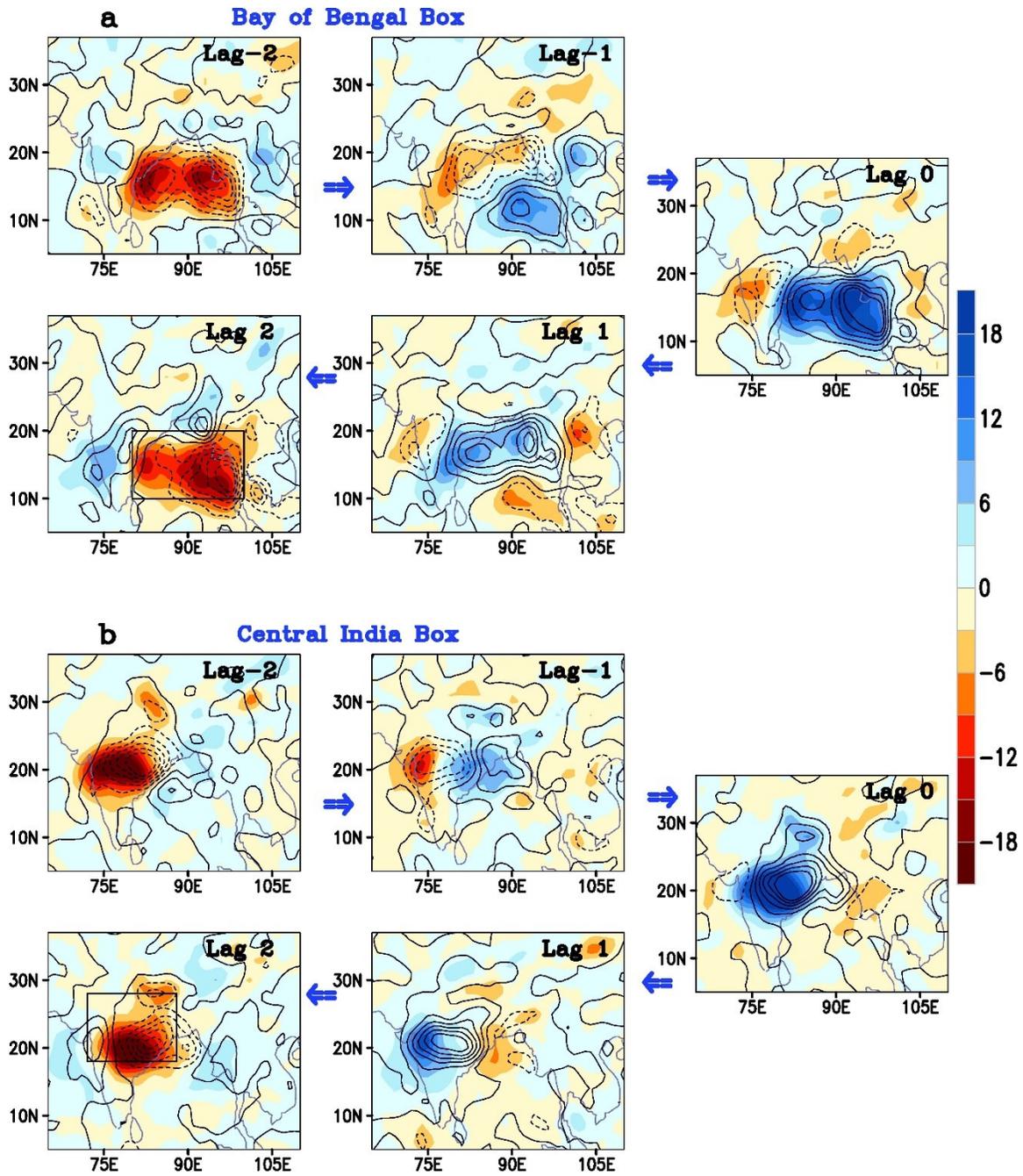

**Figure 13:** Same as Figure 11, but for for CFSv2 model experiments (MPHY).

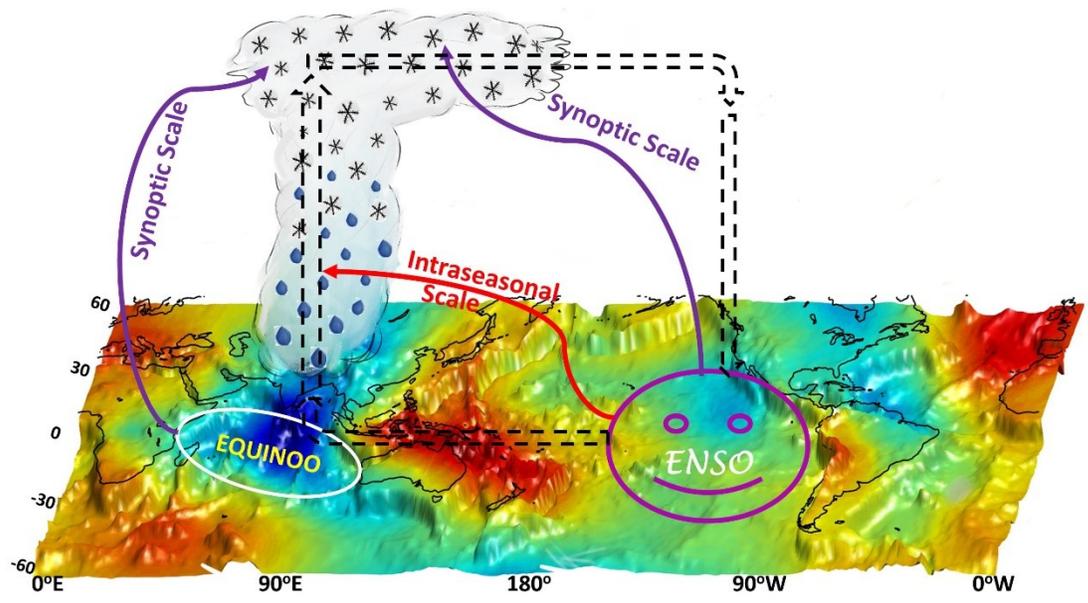

**Figure 14:** Schematic of association of Cloud condensates with Large scale process